\documentclass[conference]{IEEEtran}
\IEEEoverridecommandlockouts
\usepackage{cite}
\usepackage{amsmath,amssymb,amsfonts}
\usepackage{algorithmic}
\usepackage{graphicx}
\usepackage{textcomp}
\usepackage{xcolor}
\usepackage{url}

\def\BibTeX{{\rm B\kern-.05em{\sc i\kern-.025em b}\kern-.08em
    T\kern-.1667em\lower.7ex\hbox{E}\kern-.125emX}}
\begin{document}

\title{Resource- and Message Size-Aware Scheduling of Stream Processing at the Edge with application to Realtime Microscopy
\thanks{
This work is funded by the Swedish Foundation for Strategic Research (SSF) under award no. BD15-0008, and the eSSENCE strategic collaboration on eScience.
}
}


\author{
    \IEEEauthorblockN{
    Ben Blamey\IEEEauthorrefmark{1},
    Ida-Maria Sintorn\IEEEauthorrefmark{1}\IEEEauthorrefmark{2},
    Andreas Hellander\IEEEauthorrefmark{1},
    and Salman Toor\IEEEauthorrefmark{1}
    }
    \IEEEauthorblockA{
    \IEEEauthorrefmark{1} Department of Information Technology, Uppsala University, Sweden\\ 
    Email: \{Ben.Blamey, Ida.Sintorn, Andreas.Hellander, Salman.Toor\}@it.uu.se\\
    \IEEEauthorrefmark{2} Vironova AB, Stockholm, Sweden \\
    Email: ida.sintorn@vironova.com
    }
}

\maketitle

\thispagestyle{plain}
\pagestyle{plain}

\begin{abstract}
Whilst computational resources at the cloud edge can be leveraged to improve latency and reduce the costs of cloud services for a wide variety mobile, web, and IoT applications;  such resources are naturally constrained. For distributed stream processing applications, there are clear advantages to offloading some processing work to the cloud edge. Many state of the art stream processing applications such as Flink and Spark Streaming, being designed to run exclusively in the cloud, are a poor fit for such hybrid edge/cloud deployment settings, not least because their schedulers take limited consideration of the heterogeneous hardware in such deployments. In particular, their schedulers broadly assume a homogeneous network topology (aside from data locality consideration in, e.g., HDFS/Spark). Specialized stream processing frameworks intended for such hybrid deployment scenarios, especially IoT applications, allow developers to manually allocate specific operators in the pipeline to nodes at the cloud edge. In this paper, we investigate scheduling stream processing in hybrid cloud/edge deployment settings with sensitivity to CPU costs and message size, with the aim of maximizing throughput with respect to limited edge resources. We demonstrate real-time edge processing of a stream of electron microscopy images, and measure a consistent reduction in end-to-end latency under our approach versus a resource-agnostic baseline scheduler, under benchmarking. 
\end{abstract}


\section{Introduction}
The key idea evaluated in this paper is that prioritizing messages for processing by a stream operator at the edge; according to the estimated extent of message size reduction under that operator; is a means of making most effective use of compute resource at the edge, in cases where cloud upload speed is a bound on overall stream processing throughput.

The general stream processing pipeline considered is as follows: a stream of documents originate the cloud edge, where they can be optionally processed, individually, by some stream operator. This stream operator will transform a message thereby reducing its size, with some per-message CPU cost. This has the consequence of reducing the upload time required for that message. The stream of documents, whether processed or not, are then streamed into the cloud for further processing. We suppose that such a stream operator utilizes a single processing core, and that the edge node has several cores, so that several messages can be processed in parallel. The messages are uploaded concurrently, subject to an overall bound on network throughput.

Network (or indeed, Internet) upload bandwidth can be a bottleneck on overall throughput for such a system, if documents are sufficiently large, and arrive sufficiently frequently. Similarly, depending on the CPU costs associated with processing these documents, there may be insufficient CPU resources at the cloud edge to process all the documents without impacting overall throughput. If our goal is to maximize overall throughput, we should utilize any spare CPU capacity at the cloud edge, to reduce the size of documents queued for upload.

The key idea of this paper is whether the structure of the stream can be exploited to inform scheduling decisions about which documents should be processed at the edge, using metadata available from the stream (such as message size, relative position, and so forth). The goal of such a scheduler is to select messages for processing at the cloud edge, such that the reduction in message size, is maximized, as this will yield the greatest possible throughput. Note that throughout the paper, all mentions of \emph{reduction in message size}, \emph{ratio} (and similar), should be read as meaning this ratio divided by associated CPU cost of processing that message.

We evaluate this idea in the context of a use case based on real time analysis of images from a microscope. In our case study, the processing operator is an application-specific image de-noising operation, which has the effect of removing noise from particular areas of the image, hence reducing the image size under lossless PNG compression. This processing operator merely serves to demonstrate the concept, and is not of particular interest to the study. 

The wider application of this work concerns quasi real-time control of microscopy instrumentation, driven by image analysis performed in the cloud. This broader application is outside the scope of this paper. 

\section{Background and Motivation}

This study is motivated by a real time streaming of transmission electron microscopy images. The human operator typically begins with a pre-scan of the sample, which is then reviewed to find areas of interest, then manually targeted for further, higher magnification (zoomed-in) imaging.

A wider goal of our project is to automate aspects of the human work involved in performing this sample analysis with electron microscopes. For this paper, use an example dataset captured from such an initial pre-scan. We are working on developing a machine-learning based pipeline for the analysis of these images in the cloud, which will output instructions for the microscope to follow in subsequent phase of more zoomed in scanning. This larger pipeline is outside the scope of this paper, our focus here is on the upload of the initial image stream from the cloud edge into the cloud. 

More broadly, uploading microscopy images to the cloud has a number of potential benefits:

\begin{itemize}
\item Cloud resources can be used for processing and storing images; with low up-front cost, allowing you to pay only for what you use. Cloud providers can exploit economy of scale to offer services at low cost. Cloud providers manage the maintenance, backup, etc., according to the infrastructure-as-a-service business model.

\item For storage in particular, a variety of different storage grades are available, from high-cost high-performance storage, to low cost archive storage. Additionally, Specialist hardware such as GPUs can be rented for machine learning tasks, for example.

\item Uploading images into the cloud means they can be accessed by experts who are geographically distributed, for inspection, review, etc.

\item Cloud-based applications can be scaled up according to demand, and scaled down when idle, so that resources do not need to be paid for all the time.

\end{itemize}

A desktop PC is connected to the microscope, running vendor-specific software which is able to stream images to the disk. The \emph{HASTE Desktop Agent} stream processing tool presented in this paper monitors the target directory, and queues images for processing, and upload to the cloud. The processing function described in Section~\ref{sec:experimental_setup} reduces the size of the messages. But, as is typical in edge computing scenarios, compute resources at the edge are limited, and not all images can be processed to maintain pace with the upload. Scheduling this processing and upload at the cloud edge, for effective use of resources for both compute and upload at the cloud edge, is the focus of this paper. Our approach is described in Section~\ref{sec:methodology}.



\section{Related Work}

There is currently extensive interest in edge computing, i.e. geographically distributed provision of computational resources outside traditional datacenters. Such compute resources can be \emph{closer} to mobile and web clients, IoT devices and sensors. Consequently, they can reduce latencies and various costs associated with the provision of such services, as some work is offloaded to the edge, nearer the client. There are additional potential advantages concerning security and privacy~\cite{shiPromiseEdgeComputing2016}. Web caching at the edge~\cite{liuCachingWirelessEdge2016} is a simple example. It is well-known that computational resources at the cloud edge are limited by economic, power consumption and other infrastructure concerns; in comparison to cloud computing resources. Consequently, effective use of these limited resources is an active research area.

The focus for this paper is stream processing, processing a stream of documents, from mobile, web or IoT applications. There are many frameworks for developing such systems, and they typically allow developers to design a pipeline of \emph{stream operators} forming a directed acyclic graph (DAG), through which data flows and is processed. Often, these documents are small \emph{tuples}, representing a JSON or XML document, or perhaps a single line in a log file. 
Ubiquitous enterprise stream processing frameworks such as Apache Spark Streaming~\cite{maaralaLowLatencyAnalytics2015}, Flink~\cite{carboneApacheFlinkStream2015}, are intended for deployment in on homogeneous cloud hardware naturally take little consideration in their scheduling on heterogenous network hardware. Spark has support for data-locality – it will try to schedule analysis tasks where data resides (for example, the local HDFS data node). Spark has a simple scheduling heuristic for data-local operations: by default, tasks will execute on the same node as the data, but after a timeout, the work will be scheduled remotely (configured by the \texttt{spark.locality.wait}, and related, settings). Intended to run at large scale, elasticity and scalability are key design concerns for these frameworks. They are intended to handle message frequency throughput towards MHz, and pipelines of arbitrary complexity. Since the individual messages are small, resource management is built around monitoring the resource usage and tuple throughput, and latency~\cite{heinzeLatencyawareElasticScaling2014} of the operators -- and scaling them accordingly.

There is extensive interest in developing stream processing platforms intended for deployment in cloud/edge contexts. Scheduling is a key challenge for such systems: broadly, deciding which messages should be processed (and when), through which stream operators in the pipeline, and at which processing nodes in the cloud and the edge. Pushing operators to the edge is challenging due to resource restrictions (CPU, memory, bandwidth). The research problem is often framed in terms of assigning stream operators to nodes in a distributed system -- the so called \emph{operator placement} problem -- to maximize the performance according to some metric (such as maximum throughput, or latency requirements). 

Benoit et al.~\cite{benoitSchedulingLinearChain2013a} have shown that (under some formulations) operator placement on heterogenous hardware is NP-Hard. One approach is to allow developers to specify where their operators should be placed, though code annotations or rules. R-Pulsar \cite{renartEdgebasedFrameworkEnabling2019} is such a framework, intended for IoT applications. SpanEdge \cite{sajjadSpanEdgeUnifyingStream2016} is a similar framework, allowing developers to specify which operators need to be close to the streaming source. Their scheduler handles operator placement according to the developers' wishes, with the goal of minimizing latency.

Other work explores various strategies and heuristics for automating scheduling the stream processing, mainly concerned with operator placement. \cite{dongJointOptimizationTask2019} use an evolutionary game theory approach to decide how to offload work from the cloud onto mobile edge devices, with a somewhat complex model of multiple users, edge clouds, tasks, costs, and payoff functions. Focusing on mobile devices, power consumption is a key consideration of their work, unlike our context, where the edge device (i.e. a desktop PC connected to the web) is mains-powered. 

Whilst not focused on stream processing specifically, \cite{fahsProximityAwareTrafficRouting2019} consider routing traffic based on network proximity metrics. Their focus is on latency – our study by contrast concerns making full use of the internet uplink, where it is a throughput bottleneck for the overall system. \cite{chavesSchedulingCloudApplications2013} consider scheduling under uncertain bandwidth estimates, whereas \cite{genezEstimationAvailableBandwidth2019} consider the impact of bandwidth estimation on scheduling, in the context of intra-cloud communication.


Our approach is to consider a single stream operator. In scientific computing and microscopy contexts, execution cost can be dominated by a single operation~\cite{benblameyApacheSparkStreaming2019}, perhaps extracting features from a large document (like an image), followed by subsequent analysis on those extracted features. The case study in this paper is typical of such a scenario, the input documents are sufficiently large, and their processing sufficiently costly, that there are insufficient compute resources at the cloud edge to process all the documents in real time. We frame scheduling as \emph{message prioritization} rather than \emph{operator placement}. To the authors' knowledge, no existing work explores our scheduling approach, driven by per-message estimation of resource usage and message size reduction.

\section{Methodology}
\label{sec:methodology}
For this study, we implemented a scheduler which prioritizes documents according to the expected extent to which their message size is reduced under the processing operator, normalized with respect to the CPU cost of that operator for a single message. 

In this section, we discuss the prioritization-based approach, the means of predicting this reduction ratio with linear splines, and the means of evaluation. Details of the dataset, choice (and implementation) of processing operator, and deployment used for the evaluation, all of secondary concern to the main argument, are given separately in Section~\ref{sec:experimental_setup}. 

\begin{figure}
    \centering
    \includegraphics[width=\columnwidth]{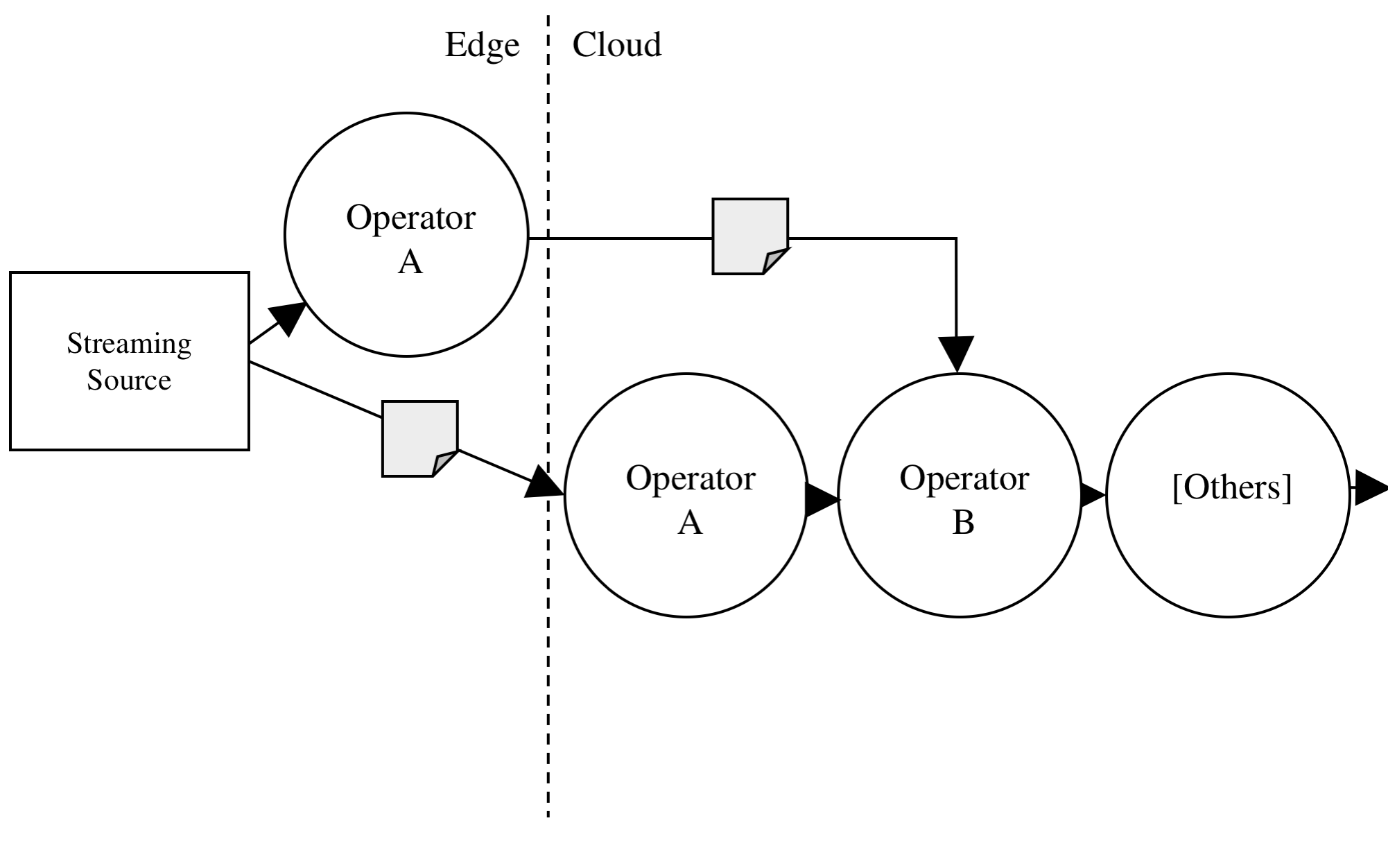}
    \caption{Logical Stream Processing Pipeline. We consider the execution of Stream Operator \emph{A} - which can be executed on a document either at the edge or the cloud. This paper concerns the (optimal) selection/prioritization of documents for such edge processing, and which to upload without processing.}
    \label{fig:pipeline}
\end{figure}

To demonstrate key idea (scheduling with consideration of the CPU-normalized reduction in message size), we consider a simple stream pipeline (Figure~\ref{fig:pipeline}), with a single stream processing operator, with other operations performed in the cloud. We consider only message prioritization at the edge, as part of a wider system. We consider a map operator, which inputs a single document, and outputs a single document. Documents (images) arriving at the stream edge, may, or may not be processed by this operator at the edge (according to the scheduler), and are then uploaded to the cloud in either case, for continued processing in the pipeline. 

Our goal is to maximize throughput, which we measure as end-to-end latency to process a fixed number of documents (see results in Figure~\ref{fig:boxwhisker}). That is, the time elapsed from the arrival of the first document, to the completion of the upload of the final document. We measure the performance of our approach against a baseline where documents are processed in arrival order for processing at the cloud edge. The goal is to fully utilize the available upload bandwidth, whilst utilizing any available CPU resource to process messages to reduce the message size, and hence improve overall pipeline throughput.

\subsection{The Scheduler}
This section describes the key ideas in the software, deferring implementation details to Section~\ref{sec:experimental_setup}.

\begin{figure}
    \centering
    \includegraphics[width=\columnwidth]{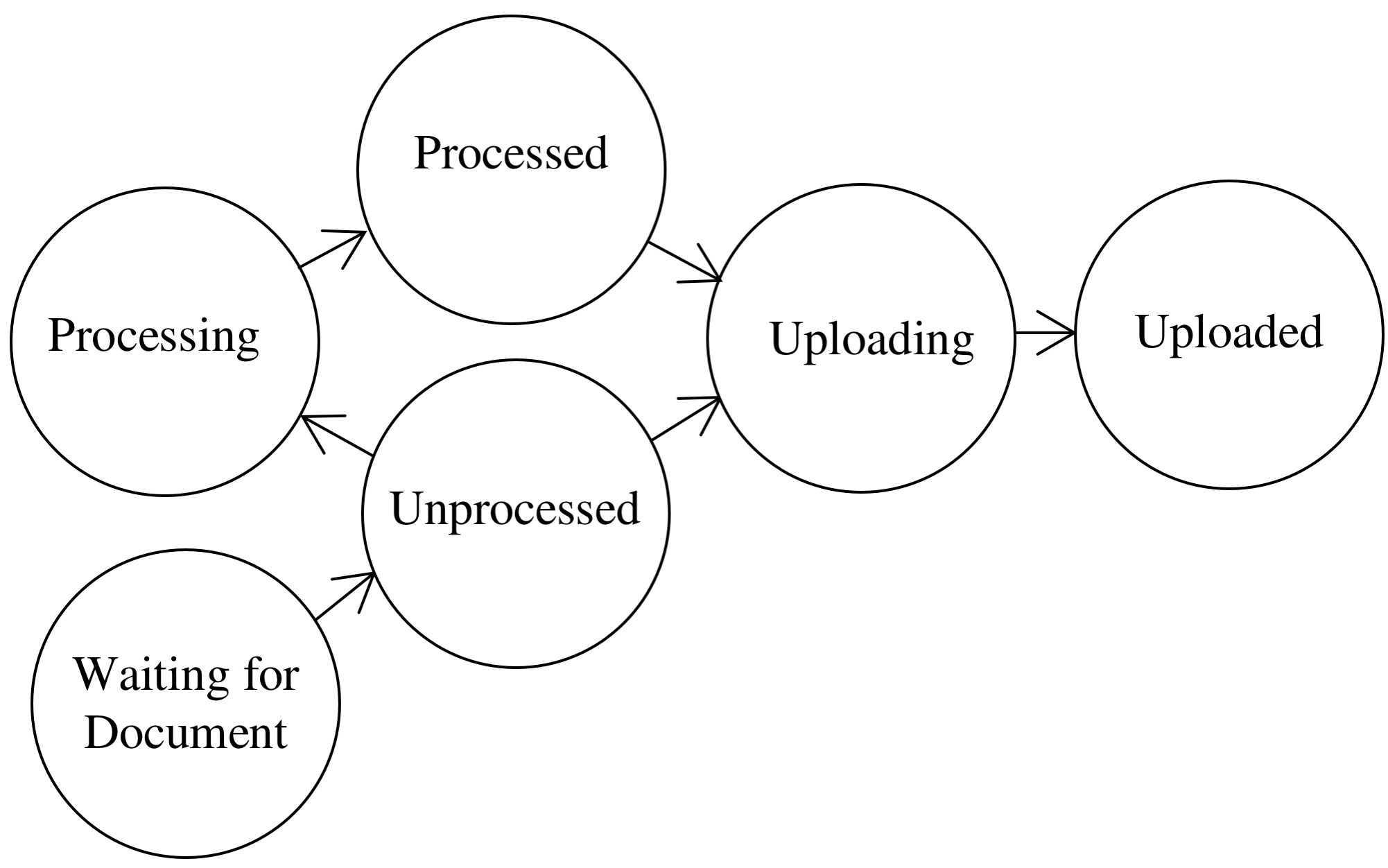}
    \caption{State transition diagram for document life cycle. `Processing' (etc.) means processed by the stream operator, and `Uploaded' means uploaded into the cloud. Each document is always at exactly one of these states at the cloud edge. But different documents are concurrently arriving, being processed, and being uploaded -- i.e. transitioning among these states.}
    \label{fig:states}
\end{figure}

The application holds a queue of messages waiting to be uploaded to the cloud. New messages are added to this queue (in this case, as they are captured by the microscope). Messages are uploaded from the queue into the cloud via a REST API. The user is able to configure the limit on the number of concurrent uploads, to make full use of available bandwidth. Additionally, documents (images) can be processed by the stream operator (again the user can configure the degree of parallelism), and then re-added to the queue, having been processed. Messages that are being processed cannot be uploaded, and vice-versa. Images that have been uploaded are no longer available for processing. The life cycle of a message at the cloud edge is visualized as a state transition diagram, Figure~\ref{fig:states}.

When stream processing is underway, there are, say, N concurrent uploads, and M concurrent `slots' for processing messages at the cloud, and a backlog of documents waiting in the queue at the edge (supposing that the ingress rate exceeds the upload rate as determined by the upload bandwidth), some mixture of processed and unprocessed documents. When a document is finished being processed or uploaded, the scheduler must make a decision about which document to process (or upload next), in order to maximize throughput. That scheduling decision is the focus of this paper: our approach is prioritize according to the extent of estimated reduction in message size (normalized by CPU cost) under the stream operator, described below. 

\subsection{Estimation of CPU-Normalized Message Size Reduction}
\label{subsec:ratio}
We assume that the first processing operator will reduce the size of the message by some amount of bytes. This will incur some CPU cost, and at an edge node, compute resources are likely to be limited~\cite{dongJointOptimizationTask2019}. Depending on the various factors, it is possible that there is insufficient processing resource to process all messages prior to upload. If our goal is to maximize overall throughput, it makes sense to prioritize messages for processing where the reduction in message size (normalized by CPU cost) is greatest, supposing that upload time is (approximately) linear in cumulative message size. Consequently, it makes sense to adopt the inverse policy for document upload: prioritizing the upload of documents where CPU resource would be \emph{least} effectively applied -- and deferring processing these messages to the cloud, where there much greater compute resources are available. Its also clear that processed documents should have higher upload priority than unprocessed documents. The spline estimation can be locally re-calculated each time a document finishes pre-processing; and the queue re-ordered according to the revised estimates of message size reduction.

So, the challenge is how to estimate this message size reduction ratio, for new data, during stream processing, where the ratio is unknown \emph{a priori}. As discussed in Section~\ref{sec:experimental_setup}, our particular choice of input data, and choice of processing operator, exhibit a phenomenon where documents located nearby one another in the stream (that is, have neigboring stream index), and are likely to exhibit similar reduction ratios. It is this phenomenon that we exploit for scheduling. 
On this basis, by measuring the elapsed CPU time (and message size reduction), for those documents that are processed at the edge, we can estimate the ratio for other documents. In practice, for this case study, the ratio is an irregular function of document index (see Figure~\ref{fig:splines}). We chose to use linear splines as a robust and simple means of estimating the ratio. This estimates the ratio based on the outcome of neighboring documents. Linear splines are also cheap, which is beneficial given that these calculations need to be made at low latency, at the edge node, where compute resources are limited.

Finally, a sampling strategy is required, to balance the exploitation of regions of the stream found to exhibit a high degree of message size reduction, with the competing need to discover new regions of high and low message size reduction. The simple strategy we propose is to select a message from an `unknown' region of the stream, for every 5th message to process -- which felt like a reasonable balance. 






These heuristics were implemented within the \emph{HASTE Agent}, together with the \emph{HASTE gateway} deployed in the cloud. The performance of this system was then benchmarked with the scheduling heuristics described above, together with various control and baseline configurations, including implementation where documents were selected for processing and upload in random order.

\section{Experimental Setup}
\label{sec:experimental_setup}

\subsection{Dataset and Stream Operator}
The dataset consists of 759 8-bit greyscale PNG images, captured using MiniTEM\textsuperscript{TM} - a 25keV transmission electron microscope (Vironova, Sweden).

\begin{figure}
    \centering
    \includegraphics[width=0.6\columnwidth]{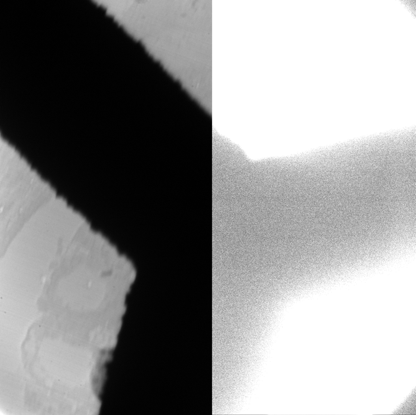}
    \caption{Image from the MiniTEM showing noise in obscured areas. Left section: original image; sample is visible in the grey areas, dark areas are the honeycomb grid obscuring the sample. Right: artificially increased contrast and brightness to illustrate noise in region obscured by the grid. Removal of noise from these areas with the processing operator can significantly reduce image size under lossless PNG compression.}
    \label{fig:noise}
\end{figure}

Much of the sample is obscured by a honeycomb shaped grid which supports the physical sample (in this case a thin tissue section). These darker obscured areas do not contain useful information about the sample. However, whilst dark, these regions are noisy (see Figure~\ref{fig:noise}), meaning little (lossless) compression is possible. Replacing the regions with uniform black pixels greatly improves their compressibility, resulting in a reduced PNG image size. 

For some images in the sequence, much of the image is obscured by the grid, so a reduction in file size of up to 40\% is possible, whilst leaving regions of the image where the sample is visible unchanged (see Figure~\ref{fig:noise}). For other images, there is little or no grid visible, so the file size reduction is minimal. Since there is overhead in opening and modifying the images, there is variance in the effectiveness of the image (i.e. message) size reduction when normalized by CPU cost. 

Crucially, since the images are taken in sequence as the instrument moves over the sample, there is a relationship between the index of the document, and the extent to which the grid is visible, and so the extent of message size reduction. Hence, the effectiveness of the operator is an irregular function of document index, as shown in Figure~\ref{fig:splines}. Whilst unknown \emph{a priori}, this relationship can be leveraged by the scheduler to prioritize documents for processing, through online estimation of the message size reduction. The stream processing operator serves to demonstrate the overall concept, and not intended as a scientific contribution in its own right.

The stream processing operator used for this application is simple:
\begin{enumerate}
\item Surround the image with a black border of width 1 pixel.
\item ‘Threshold fill’ the image with black, using the so called `forest-fire' implementation of flood fill.
\item Crop the 1 pixel border.
\end{enumerate}

The processing operator was implemented in Python. The threshold intensity for the fill was 30.

\subsection{Implementation}

\begin{figure}
    \centering
    \includegraphics[width=\columnwidth]{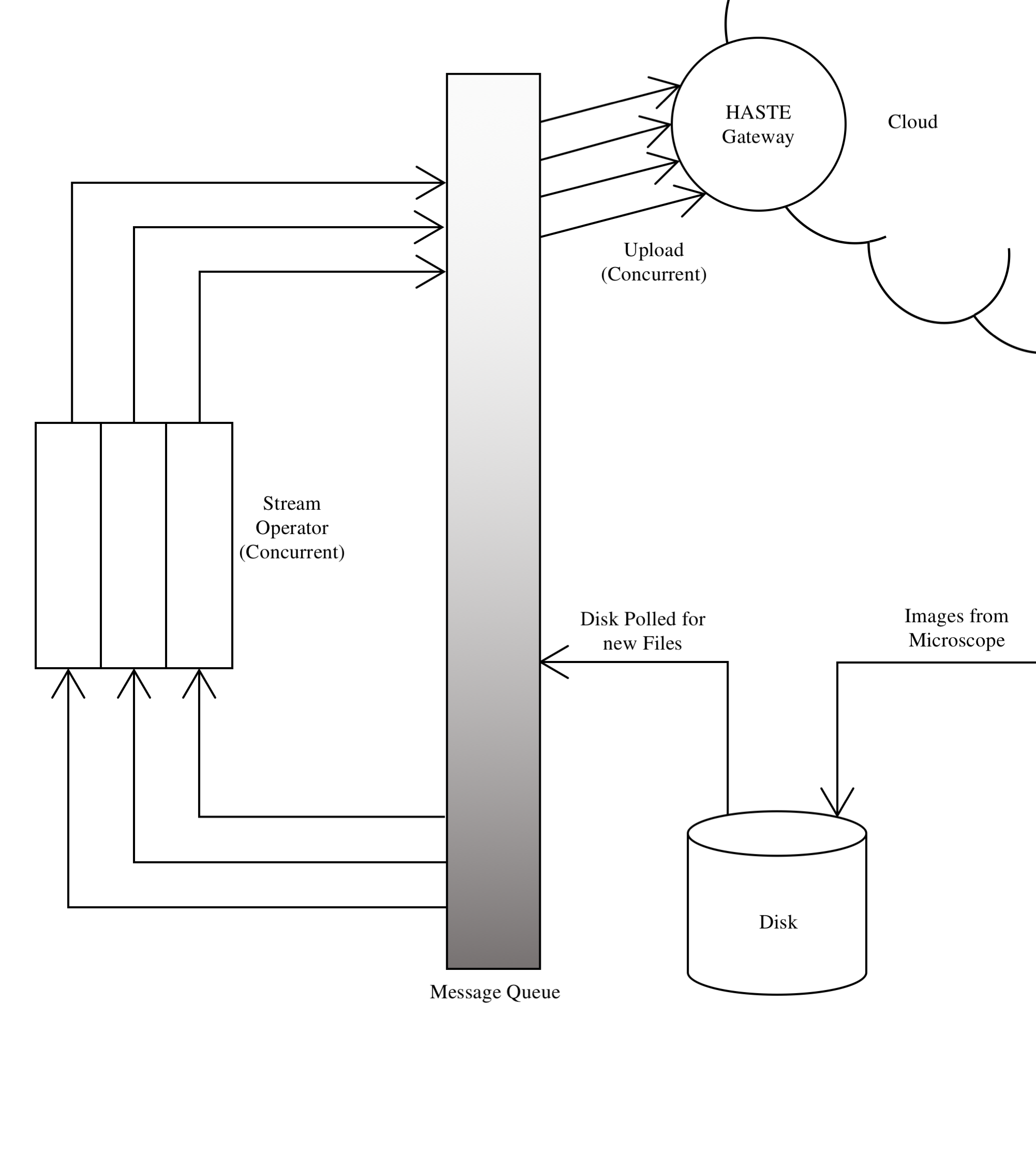}
    \caption{Overview of the implementation of the \emph{HASTE Desktop Agent}. Shows the message queue, with incoming images from disk, concurrent processing of images from the queue, and concurrent upload of images from the queue into the cloud. This is intended as a schematic, not to explain the prioritization policy associated with the message queue.}
    \label{fig:implementation}
\end{figure}

The software used in this study utilizes various components of the HASTE toolkit, open-sourced at \url{{https://github.com/HASTE-project/}}. 

\textbf{The HASTE Agent:} a desktop application which concurrently: (a) monitors the disk for new images from the microscope, (b) processes images according to the defined processing operator function, (c) uploads documents to the cloud, and (d) measures the performance of the processing function, re-computing the splines, and re-prioritizing the queued images according. This application combines both multi-threaded execution, and asynchronous method invocation (Python 3's \texttt{asyncio} functionality) to acheive concurrent message processing. (see: \url{https://github.com/HASTE-project/desktop-agent}). 
    
\textbf{The HASTE Storage Client:} library with the core priorization functionality used by the HASTE agent. (see: \url{{https://github.com/HASTE-project/HasteStorageClient}}).
    
\textbf{The HASTE Gateway:} cloud gateway service, which recieves images in the cloud. Deployed as a Docker container. Implemented with \texttt{aiohttp}. (see: \url{{https://github.com/HASTE-project/haste-gateway}}).

\subsection{Compute Resources}
The machine used as the edge streaming node for the benchmarking has single Intel Core i5, with 2 physical CPU cores (4 virtual cores with hyperthreading). The HTTP Server which receives in the files in the cloud runs inside a Docker container on its own VM with 1 vCPU and 512Mb RAM. The Internet bandwidth was capped at 100 Mbps download and 16 Mbps upload during the benchmarking, to ensure consistent speed across runs. The cloud computing  resources were provided by SNIC~\cite{ssc}.

\section{Results}


\begin{figure}
    \centering
    \includegraphics[width=\columnwidth]{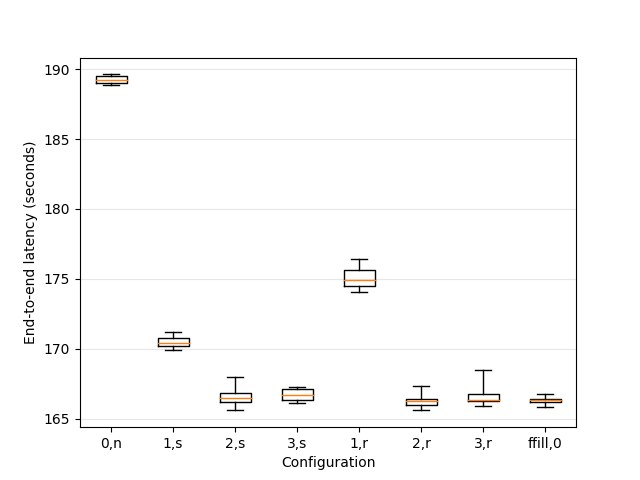}
    \caption{
    End-to-end processing time, under various configurations. Processing time is from arrival first image, to completion of upload of the last. Configurations are listed in Table~\ref{table:configs}.}
    \label{fig:boxwhisker}
\end{figure}

\begin{table}
\centering
\begin{tabular}{ |c|p{6cm}| } 
\hline
Key & Description of Configuration \\ \hline \hline
\texttt{0,r} & control: no processing, original images uploaded.\\ \hline
\texttt{1,s} & processing with 1 core, splines-based sampling.\\ \hline
\texttt{2,s} & processing with 2 cores, splines-based sampling.\\ \hline
\texttt{3,s} & processing with 3 cores, splines-based sampling.\\ \hline

\texttt{1,r} & processing with 1 core, random ordering (baseline).\\ \hline
\texttt{2,r} & processing with 2 cores, random ordering (baseline).\\ \hline
\texttt{3,r} & processing with 3 cores, random ordering (baseline).\\ \hline

\texttt{ffill,0} & control: source images pre-processed offline before starting the stream.\\ \hline
\end{tabular}
\caption{Key for configurations used for the benchmarking results presented in Figure~\ref{fig:boxwhisker}}
\label{table:configs}
\end{table}

\begin{figure}
    \centering
    \includegraphics[width=\columnwidth]{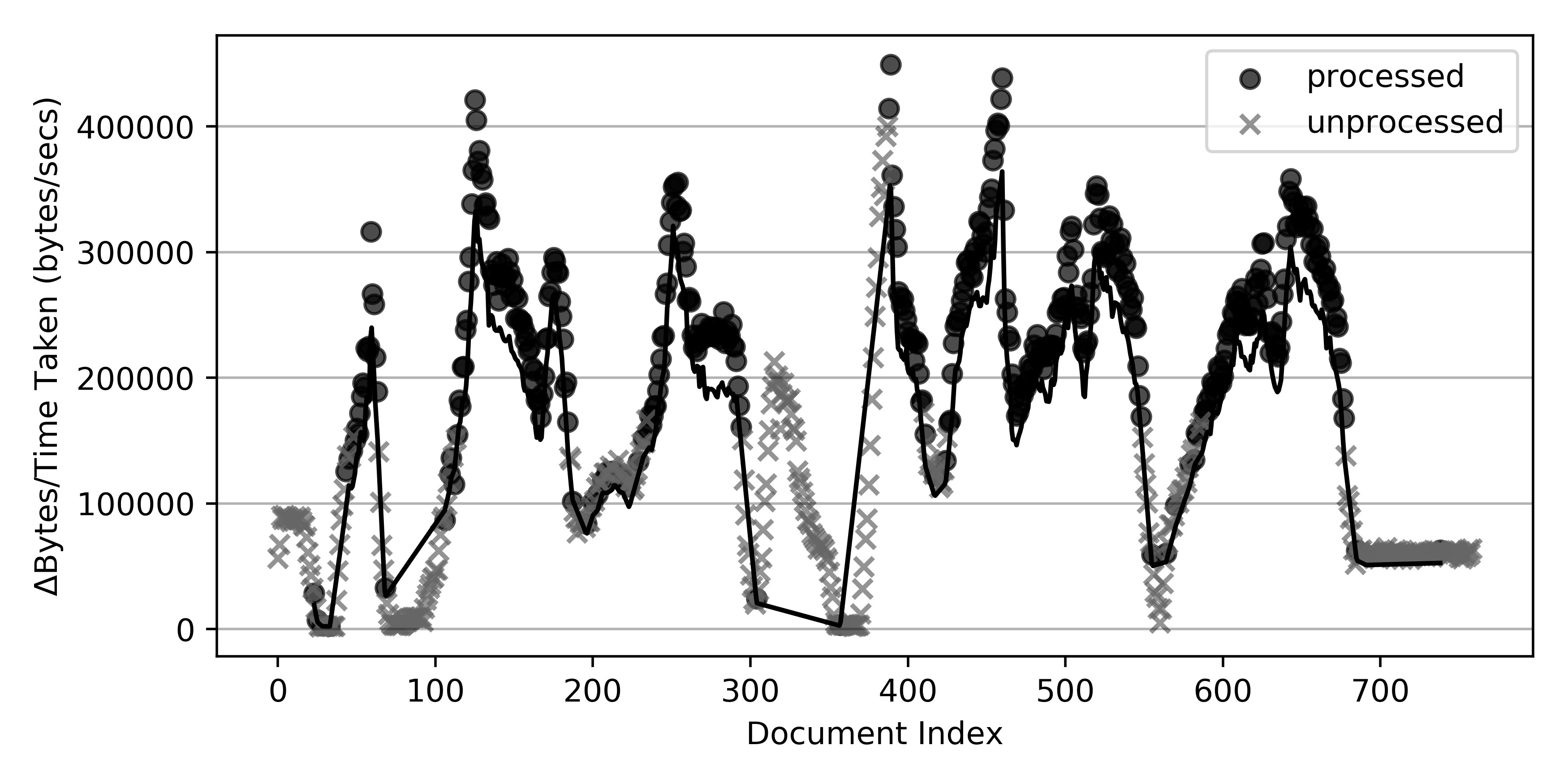}
    \caption{
    Message size reduction (normalized by CPU cost) over document index, showing which documents processed at the edge, for a single run under configuration \texttt{(1,s)}. Documents marked `processed' were processed at the cloud edge prior to upload (and vice-versa). The line shows the final revision of the splines estimation of the message size reduction. Note how this deviates from the true value (measured independently for illustration purposes on the same hardware), in regions of low reduction.}
\label{fig:splines}
\end{figure}

\begin{figure}
    \centering
    \includegraphics[width=\columnwidth]{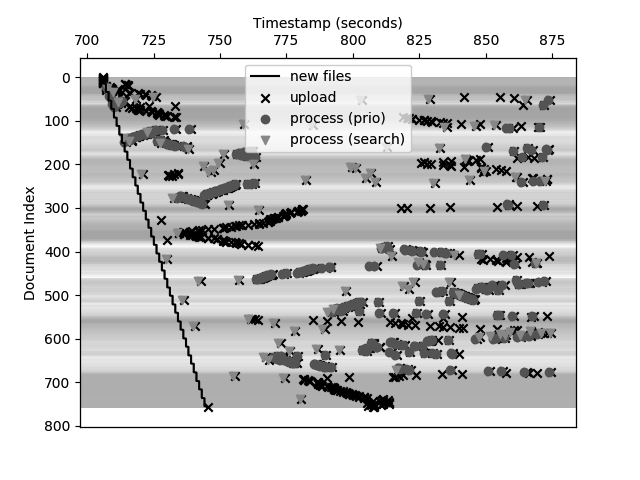}
    \caption{
Visualization of an example log trace under configuration \texttt{(1,s)}. The red line shows the arrival time of new images from the microscope. Points labelled `upload' show when an image is uploaded to the cloud. Points labelled `process (prio)' are selected for processing based on estimated message size reduction. Points labelled `process (search)' are selected to attempt to find new areas of high or low message size reduction. Note that the timestamp is from the beginning of the complete run (with multiple configurations), hence the non-zero start time. Background shading illustrates the bytes reduction (benchmarked offline on identical hardware for illustration purposes only).}
    \label{fig:trace}
\end{figure}

The system was benchmarked under a number of configurations using the dataset, and example stream operator discussed in Section~\ref{sec:experimental_setup}. These configurations are summarized in Table~\ref{table:configs}. End-to-end latencies are shown as box and whisker plots for the various configurations (Figure~\ref{fig:boxwhisker}), the goal of our scheduling approach is to improve the throughput, and hence reduce the end-to-end latency.

The end-to-end processing time was averaged over 5 runs, for each of the configurations: a control with processing disabled (\texttt{(0,n)}), a control with all files pre-processed offline, using the splines-based method for prioritization; with 1 and 2 processes/cores for image processing, and using the random prioritization; with 1 and 2 processes/cores for image processing.

There are two control configurations, where no processing is performed at the cloud edge, and we simply measure the upload time. Under configuration \texttt{(0,r)}, and all messages are uploaded without processing by the operator. This represents an upper bound on latency. Conversely, in configuration \texttt{(ffill,0)}, all the images are processed \emph{offline}, before streaming starts, so we measure the upload time for the processed images (these having reduced size), a lower bound on latency.  

We benchmark our approach, under varying number of concurrent processing operators at the cloud edge (configurations \texttt{(*,s})). As a baseline, we perform identical processing, with the same levels of concurrency, but processing messages in random order (without our smart prioritization approach). 

In Figure~\ref{fig:boxwhisker}, we can clearly see the benefit of performing edge processing, since all other configurations show greatly reduced latency in comparison to not performing any edge processing at all. But our key finding is that when we are constrained to a single concurrent processing operator at the cloud edge, we see that our approach offers an improvement over baseline random prioritization for processing, a consistent reduction in latency of \textasciitilde5 seconds. Clearly, the magnitude of this latency reduction is determined by the various factors of the application, and is not a direct measure of the effectiveness of the scheduling (see `Sensitivity', Section~\ref{sec:conc}). Indeed, the parameters are such that when the number of cores at the edge allocated to the stream operator is increased to 2 or 3, we see little advantage in our approach over the baseline (\texttt{(2,r)}, \texttt{(3,r)}) -- this is because at this level of concurrency, the processing capacity at the edge is sufficient to process all (or nearly all) the incoming documents -- and their ceases to be any benefit to prioritization. In these cases, the end-to-end latency is similar to the case where the processing is performed offline \emph{a priori}, as shown in configuration \texttt{(ffill,0)}. Again, this is a consequence of the particular performance factors for this application case, not a weakness in the scheduling approach itself.


Figure~\ref{fig:splines} visualizes the splines-based approach to message scheduling prioritization. It shows the reduction in size for each message (i.e. image), normalized by CPU cost. We can clearly see the variability in the extent of message size reduction under the operator, the motivation for our prioritization approach.
The goal of the system is to prioritize messages for processing where this reduction is the greatest (leading to the greatest reduction in upload time, assuming upload time is a linear function of document size). It does this without \emph{a priori} knowledge of the extent of this reduction -- instead estimating it online by measuring the performance characteristics (that is, message reduction) of the operator, together with the sampling strategy described in Section~\ref{subsec:ratio}. The dots labelled `processed' were processed at the edge, we can clearly see that the sampling strategy was able to effectively select messages for edge processing which exhibited higher reduction (the y-axis). The line shows the final revision of the splines estimate, which as expected, deviates from the actual message reduction, especially where documents were not processed.

Figure~\ref{fig:trace} gives a visualization of the activity of the system during processing, and the sampling strategy, as new images arrive, and concurrently move through the state transition diagram shown in Figure~\ref{fig:states}. The dots in the figure effectively represent the timestamps (x-axis) for selected transitions within this state diagram for particular documents (indexed on the y-axis). We distinguish between documents which were selected for edge processing according to the sampling heuristic: those selected for the purposes of discovering the extent of reduction in new regions (\texttt{process (search)}), and those selected to exploit known regions of `known' high reduction (\texttt{process (prio)}). We clearly see `V' a shaped pattern (e.g. around document index 380), where the scheduler `climbs' up the estimated message reduction gradient, moving out from an index known to yield low reduction, in the case of upload.

\section{Conclusion}
\label{sec:conc}
This paper has investigated a means of scheduling stream processing at the cloud edge through online estimation of the extent of message size reduction (normalized by CPU cost) under a stream processing operator. The intention is to prioritize the CPU on processing messages where the message size can be reduced as much as possible, hence reducing upload times to the cloud, and overall end-to-end stream processing latency. Cloud upload over the Internet is a natural bottleneck in many IoT applications, especially those in microscopy image processing, and other scientific workloads. To the authors' knowledge, the consideration of CPU-normalized message size reduction is novel for the stream processing domain.

In our system, with a single stream operator at the cloud edge, documents are prioritized for processing where the message size reduction is expected to be the greatest relative to the CPU cost of the operation. Under this scheme, messages are uploaded to the cloud with the inverse priority: first, any documents which have been processed, followed by documents where the processing is expected to be the least effective at reducing message size. Messages for which the processing operator will yield less message size reduction, are better processed in the cloud, where more compute resources are available, allowing edge compute to prioritize message size reduction.

Message size reduction is predicted with a linear splines model, based on documents neighbouring in the stream. The assumption is that documents with neighbouring stream indices exhibit similar levels of size reduction under the stream operator. Our specific use concerns an image feed from a microscope moving over a sample, and whilst compression is ill-suited to this use case, the similarity of neighbouring images/documents has a relationship, albeit irregular, with the performance characteristics of our stream processing operator. We have shown this relationship can be leveraged for scheduling purposes. 

Our key findings are:
\begin{enumerate}
\item that offloading work to the cloud edge can significantly improve the overall throughput (measured as end-to-end latency for a fixed length stream) of cloud based image stream processing applications, under random scheduling (as a control), and measured this for an example microscopy use case. This is unsurprising.
\item that scheduling documents for edge processing with priority based on their expected CPU-normalized message size reduction further improves this throughput, and this can be measured consistently, for the application in this study.
\item that our simple sampling strategy, combined with linear splines, is a robust and effective method for estimating this reduction ratio, balancing a need to exploit regions of known `high reduction' with the search for such regions.
\end{enumerate}

Limitations and future work:

\textbf{Sensitivity:} The overall effect of such scheduling is wholly dependent on various factors in the use case: the CPU cost of processing, available cores, the message sizes (before and after processing), the incoming document frequency, network speed, and so forth. In this study, the effect was a modest, but consistent, 3\% improvement on the baseline, in terms of overall processing time. 
Dependent on these factors, one may see an order-of-magnitude increase in throughput, or no difference at all. 

\textbf{Generalizability:} Our evaluation was conducted on an image processing application, and whilst the technique (and software) is intended for more general applications, it exploits some phenomena specific to this use case.
In this case, the overhead of measuring and scheduling on a per-image basis is worthwhile. Future work could investigate how this approach could be adapted to batches of messages in more typical stream processing applications, with smaller messages. 

\textbf{Local correlation in message size reduction ratio:} Our approach exploits a phenomenon in our dataset/application where documents with neighbouring indices within the stream exhibit similar performance under the stream processing operator. A wider survey would be needed to investigate whether this relationship can be seen in other application domains; and indeed how the relationship compares in those cases. The key idea is that information readily available, (or cheaply extracted) from the stream could be used to predict the message size reduction.

In summary, we've shown that a splines-based approach to estimating the extent of message size reduction, and scheduling edge-processing accordingly, can yield higher throughput of edge/cloud stream processing systems, for applications where this reduction can be estimated from document index (or other metadata available in the stream). We claim this is a novel approach to scheduling stream processing at the edge. 
However, the extent of the benefit is entirely dependent on various performance metrics associated with the application. Whilst further work is needed to explore generalizing this approach to streams with other characteristics (such as those with smaller messages), and more complex pipelines, this work should be relevant stream processing applications in imaging (especially pipelines including compression), applications where per-message execution time is variable, including scientific computing, and other atypical stream processing applications.

\bibliographystyle{apalike}
{
\bibliography{My_Library}}

\end{document}